\documentclass[12pt]{aastex}
\begin{document}

\title{ \emph{Wide-field Infrared Survey Explorer} Observations of Young Stellar Objects in the Western Circinus Molecular Cloud}

\author{Wilson M. Liu\altaffilmark{1}}
\altaffiltext{1}{Infrared Processing and Analysis Center, California Institute of Technology, MC 100-22, 770 South Wilson Avenue, Pasadena, CA 91125, wliu@ipac.caltech.edu}
\author{Deborah L. Padgett\altaffilmark{2}}
\altaffiltext{2}{Spitzer Science Center, California Institute of Technology, MC 314-6, 1200 East California Boulevard, Pasadena, CA 91125}
\author{David Leisawitz\altaffilmark{3}}
\altaffiltext{3}{NASA Goddard Space Flight Center, Code 605, Greenbelt, MD 20771}
\author{Sergio Fajardo-Acosta\altaffilmark{2}}
\author{Xavier P. Koenig\altaffilmark{3}}

\begin{abstract}
The \emph{Wide-field Infrared Survey Explorer} has uncovered a population of young stellar objects in the Western Circinus molecular cloud.  Images show the YSOs to be clustered into two main groups that are coincident with dark filamentary structure in the nebulosity.  Analysis of photometry shows numerous Class I and II objects.  The locations of several of these objects are found to correspond to known dense cores and CO outflows.  Class I objects tend to be concentrated in dense aggregates, and Class II objects more evenly distributed throughout the region.
\end{abstract}

\section{Introduction} \label{sec-intro}
The Circinus molecular cloud is a region of active star formation with a population of young stellar objects (YSOs).  Previous studies have estimated the distance of the cloud to be several hundred parsecs \citep{rbw08}, with \citet[hereafter B99]{bally99} adopting a distance of 700 pc, with a uncertainty of up to 50\%.  Early work by \citet{vdbh} identified two embedded objects, with more recent observations uncovering H$\alpha$-emitting sources, Herbig-Haro objects, and molecular outflows \citep{rbw08,bally99,reipurth96,dob,mo94,re94,rg88}.  The region was observed at millimeter wavelengths by \citet[hereafter R96]{reipurth96} and B99.  R96 identified protostellar candidates at 1300 $\mu$m, while B99 identified large-scale CO outflows.  The Circinus region provides an good opportunity to investigate a population of young, coeval stars.

In this Letter, we present observations by the \emph{Wide-field Infrared Survey Explorer} (WISE) \citep{wright_wise} of the young cluster associated with the Western Circinus cloud.  The observations were taken in the four WISE bands (3.4, 4.6, 12, and 22 $\mu$m).  The nature of the WISE dataset makes it useful for the characterization of star forming regions.  Bands 3 and 4 (12 and 22 $\mu$m), in particular, are well-suited for the detection of warm circumstellar dust and objects embedded in dense material.  The wide field-of-view allows one to determine the YSO distribution and the structure of nebulosity over a large spatial scale.  The WISE dataset also represents a complement to the \emph{Two Micron All-Sky Survey} (2MASS), allowing one to characterize spectral energy distributions (SEDs) of YSOs in seven bands from 1.2 to 22 $\mu$m.  We summarize the observations and data processing in Section \ref{sec-obs}, present results in \S \ref{sec-results}, and discuss them in \S \ref{sec-disc}.

\section{Observations and Data Reduction} \label{sec-obs}
The WISE spacecraft conducts observations in a continuous scanning mode, using a scan mirror to freeze a single pointing on the detector.  Individual images are integrated for 8.8 seconds in all four bands.  The images are processed using the WISE Science Data System, developed at the Infrared Processing and Analysis Center, to ingest raw data, produce images, detect sources, and extract photometry.  Distortion corrections and band merging are also performed by the automated pipeline \citep{cutri11,jarrett,wright_wise,cutri09}.  Observations of the Western Circinus region were taken 2010 February 22-24.  Single images, which have a field-of-view of 47$\arcmin$ and a plate scale of 2.75$\arcsec$ per pixel, are coadded to produce images free from cosmic rays and other artifacts and a plate scale of 1.38$\arcsec$ per pixel.  The automated source extraction recovers sources detected at a specified signal-to-noise ratio which, in the case of these observations, is 5 or greater.  Positional matching of WISE sources to 2MASS point sources is automated. Typical positional offsets between WISE and 2MASS are 0.1 to 0.2$\arcsec$.  For some sources found in dense groups, band matching was completed by inspection.  

\subsection{Source selection}
Automated extraction recovers 6848 sources in a region centered near ($\alpha = 15^{h} 00^{m} 45^{s}$, $\delta = -63\degr 13\arcmin 00\arcsec$, J2000), with dimensions of approximately 90$\arcmin$ by 45$\arcmin$.  This list includes a significant amount of contamination, predominantly foreground objects and spurious extractions on nebulosity.  There are also numerous sources with poor photometry (i.e., large photometric errors or upper limit measurements).  The list was filtered to ensure photometric quality, excluding objects that have photometric errors greater than 0.2 mag or upper limits in Band 4.  There are 644 sources that meet these criteria.  Out of these, the vast majority also have Band 3 measurements with photometric errors less than 0.2 mag.  It should be noted that several of the objects with valid 22 $\mu$m  photometry but no measurement at 12 $\mu$m correspond to very red objects in the image (discussed in \S \ref{sec-veryred}).

Sources in the final list correspond predominantly to point sources in the Band 4 image, though some sources are still a result of extractions on nebulosity.  These can be filtered by omitting the lowest signal-to-noise sources; requiring SNR $\geq$ 12 is effective in eliminating contamination while leaving real sources intact.  This signal-to-noise cut roughly corresponds to a 22 $\mu$m flux density of 8 mJy.  A few spurious extractions in regions of high confusion (e.g., the Cir-MMS aggregate, see below) or nebulosity were removed by inspection.  The final source list contains 206 sources which are listed, with their photometry and spectral slope (see \S \ref{sec-class}), in Table \ref{tab-sources}.

\section{Results} \label{sec-results}
\begin{figure}
\epsscale{1.0}
\rotate
\centerline{\plotone{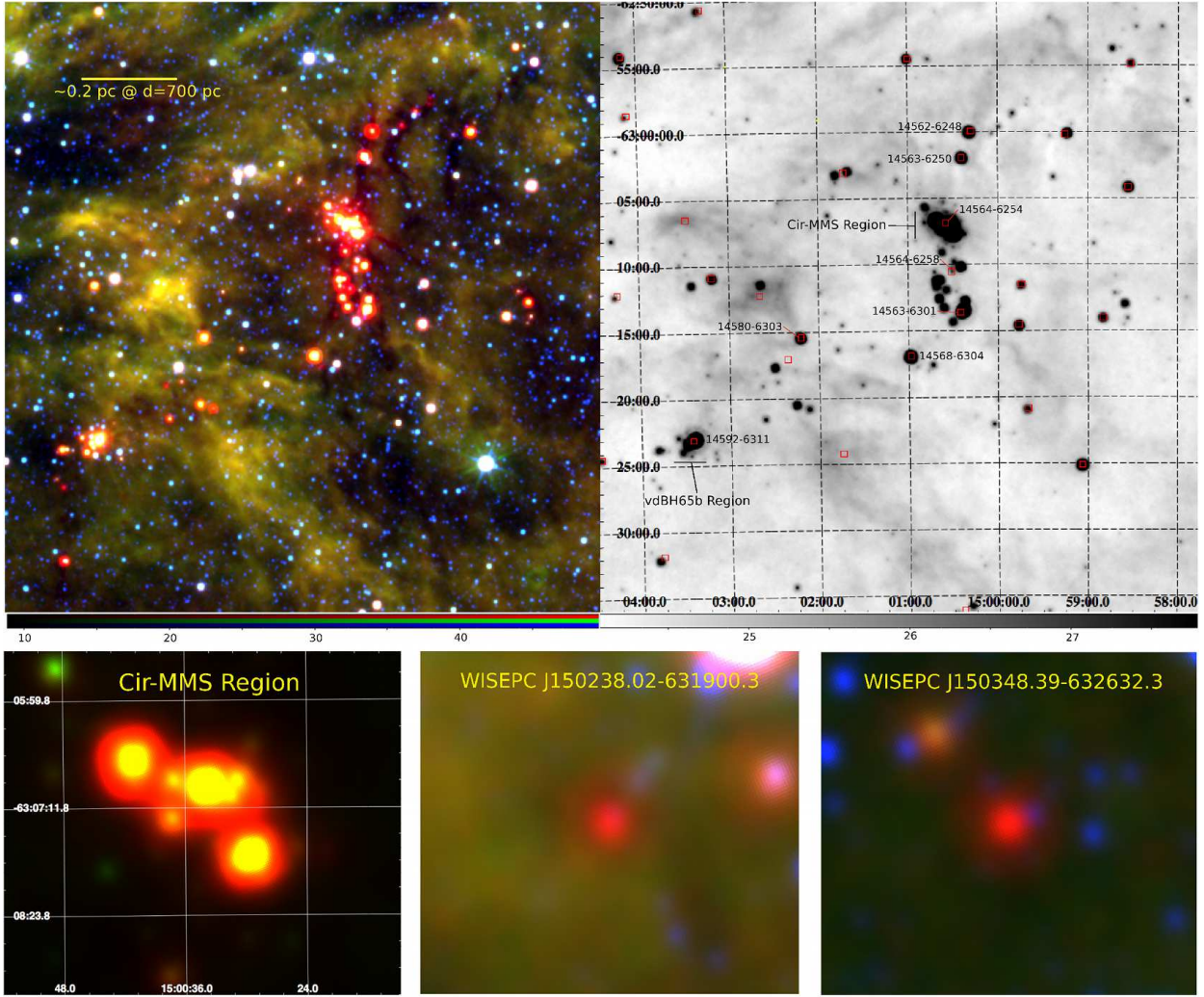}}
\caption{Top-Left: A 3-color image of the Circinus region.  22, 12, and 4.6 $\mu$m images are mapped to the the red, green, and blue channels of the image, respectively.  Top-Right: The same region showing just the 22 $\mu$m channel.  IRAS sources are noted (red boxes) and labeled with their ID for sources located along the main locus of YSOs. Bottom-Left: A 12 and 22 $\mu$m (red) image of the Cir-MMS aggregate identified by R96.  Bottom-Center and Right: Zoomed images of two of the very red objects.}
\label{fig-3color}
\end{figure}

A coadded three-color image is shown in the top-left panel of Figure \ref{fig-3color}, with a total exposure time of approximately 150 s at the image center.  The image highlights the YSO population of the region.  Several sources in the region were previously detected by the \emph{Infrared Astronomical Satellite} (IRAS) mission, and many of these correspond to individual WISE objects, as well as groups of objects.  The IRAS point sources are identified in Figure \ref{fig-3color}.  IRAS sources that correspond to single, uncrowded WISE sources are also noted in Table \ref{tab-sources}.

The YSOs in the region are clustered in two main aggregates, around the Cir-MMS millimeter sources (R96) and source 65b from \citet{vdbh}, hereafter referred to as vdBH65b.  The northern part of the cluster contains a dense aggregate of YSOs, located at $\alpha = 15^{h} 00^{m} 36^{s}$, $\delta = -63\degr 07\arcmin 00\arcsec$, coincident with a structure uncovered by R96 at 1300 $\mu$m indicating a core of dense, cold dust, and designated as Cir-MMS sources 1 through 4 (we refer to this region as the Cir-MMS core).  IRAS also shows a point source, co-located with the Cir-MMS core, 15564-6254.  The IRAS source is closest to WISEPC J150033.91-630656.6, though the large beam of IRAS would suggest that several of the sources in the aggregate contribute flux to the IRAS object.  A zoomed WISE 12 and 22 $\mu$m image of the core can be seen in Figure \ref{fig-3color}.  R96 shows an object with multiple lobes, while the WISE image shows 7 point sources detected at 12 and 22 $\mu$m.  Photometry of these objects is unreliable due to source confusion.  The estimated fluxes for the sources are noted in Table \ref{tab-sources}.

Another dense aggregate is located about 45$\arcmin$ to the southeast of the main aggregate, near $\alpha = 15^{h} 03^{m} 27^{s}$, $\delta = -63\degr 23\arcmin 21\arcsec$.  This group consists of approximately 5 sources in the 22 $\mu$m image.  The two brightest objects (WISEPC J150323.80-632258.8 and J150328.78-632316.4) lie near vdBH65b. This particular region also contains several Herbig-Haro objects, H$\alpha$ emitting stars, and other signs of active star formation \citep{rbw08,re94,mo94,rg88}. This region is discussed further in the following section.  Several objects also lie outside the main aggregates, as well as in between them.

\subsection{Photometric Characterization of YSOs} \label{sec-class}
\begin{figure}
\epsscale{0.8}
\centerline{\plotone{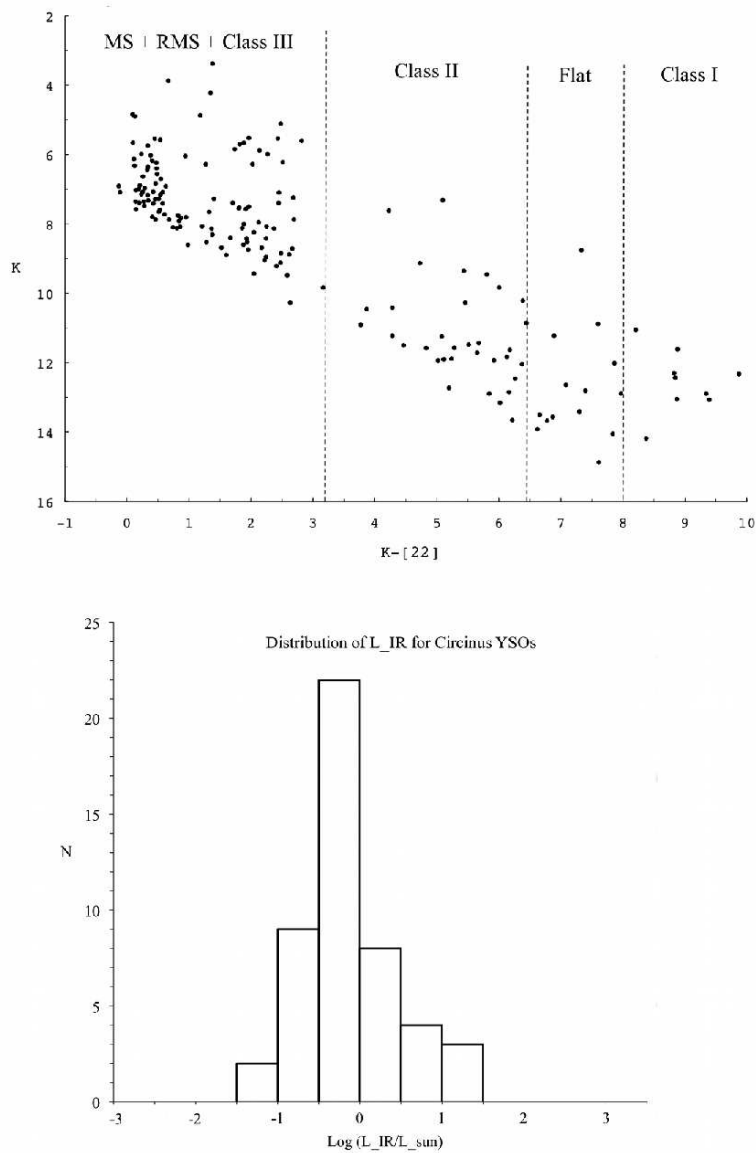}}
\caption{Top: The K$_{s}$ vs. K$_{s}$-[22] CMD for all sources with 2MASS K$_{s}$ and WISE 22 $\mu$m photometry.  The lower diagonal boundary is defined by the 22 $\mu$m sensitivity of WISE.  Contamination from extragalactic sources is not significant, except for the reddest sources at K$_{s}$ $\geq 14$.  Regions corresponding to YSO classes are shown; the boundaries are set by the $\alpha_{IR}$ ranges described in \S \ref{sec-class}, converted to K$_{s}$-[22] color.  Bottom: Distribution of infrared luminosities (1 to 26 $\mu$m) for 48 YSOs with WISE and 2MASS photometry.}
\label{fig-cmd}
\end{figure}

\begin{figure}
\epsscale{1.0}
\centerline{\plotone{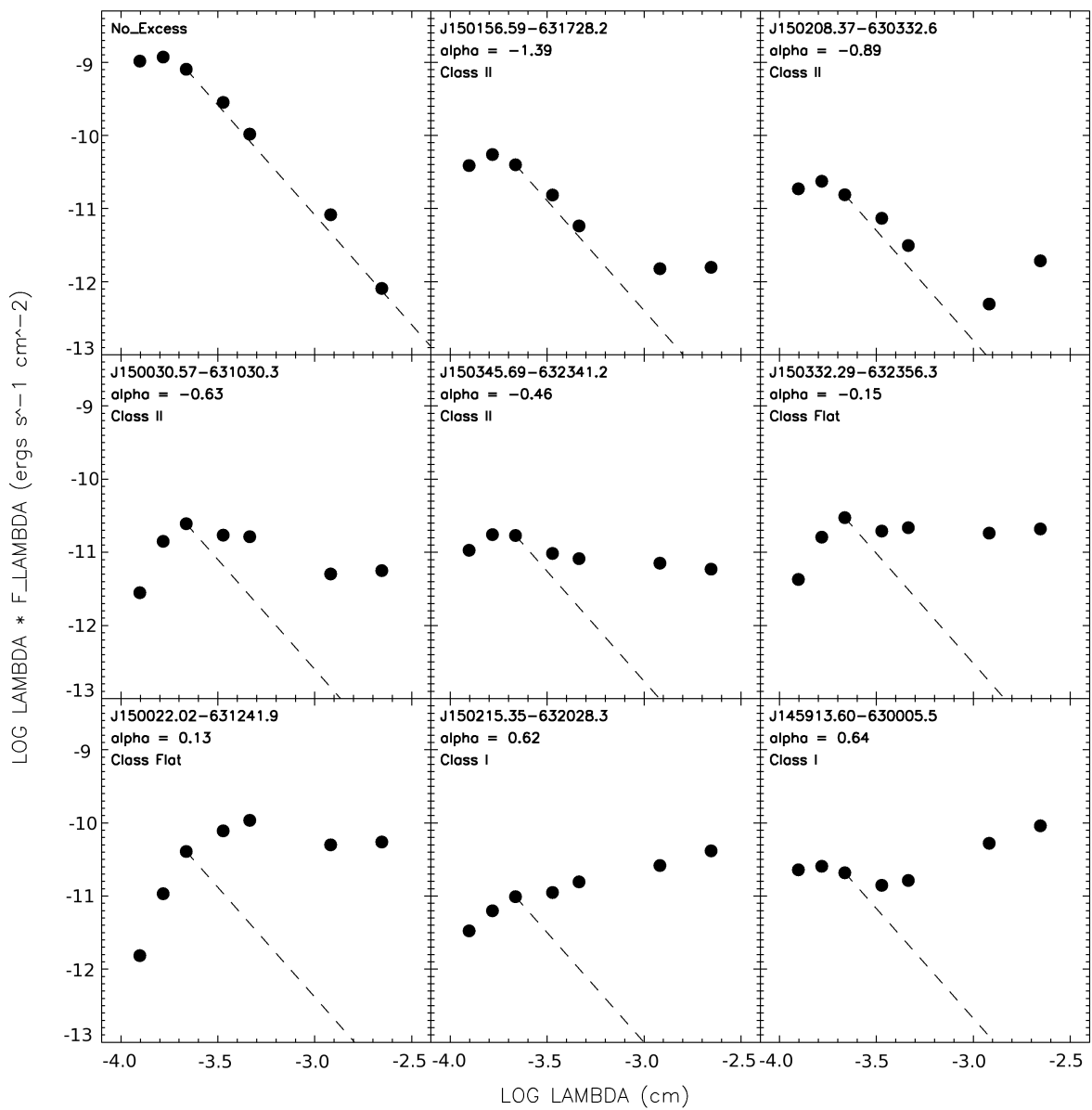}}
\caption{SEDs for eight YSOs in Circinus, with photometry in seven (WISE + 2MASS) bands.  Objects are ordered by $\alpha_{IR}$.  Object IDs are shown at the upper-left of each plot.  The dotted lines represent the Rayleigh-Jeans expectation for photospheric emission, normalized to the K$_{s}$ point.  The top-left plot shows an example of a star in the region with purely photospheric emission.  Error bars are typically smaller than the symbols.}
\label{fig-sed}
\end{figure}

For 171 objects with WISE 22 $\mu$m and 2MASS K$_{s}$ photometry, we plot a color-magnitude diagram (CMD) in Figure \ref{fig-cmd}.  For each object, we calculate the infrared spectral slope, $\alpha_{IR} = \Delta log\lambda F_{\lambda}/\Delta log\lambda$, using the 2MASS K$_{s}$ and WISE 22 $\mu$m data points.  Objects are classified into YSO classes using the criteria described in \citet{greene94}:  Class I sources have $\alpha_{IR} > 0.3$.  Flat spectrum sources have $-0.3 < \alpha_{IR} < 0.3$.  Class II sources have $\alpha_{IR} < -0.3$.  The color-magnitude regions which correspond to these classifications are noted.  For objects that do not have K$_{s}$ photometry, we use the WISE 3.4 $\mu$m point instead.  There are 21 Class I, 16 flat spectrum, and 35 Class II objects, excluding those objects with unreliable photometry.  The SEDs confirm the nature of the Class I and II objects, and show the diversity of YSO objects in the region.  Figure \ref{fig-sed} presents examples of SEDs of YSOs in the region, spanning a range of $\alpha_{IR}$.  

\begin{figure}
\epsscale{1.0}
\centerline{\plotone{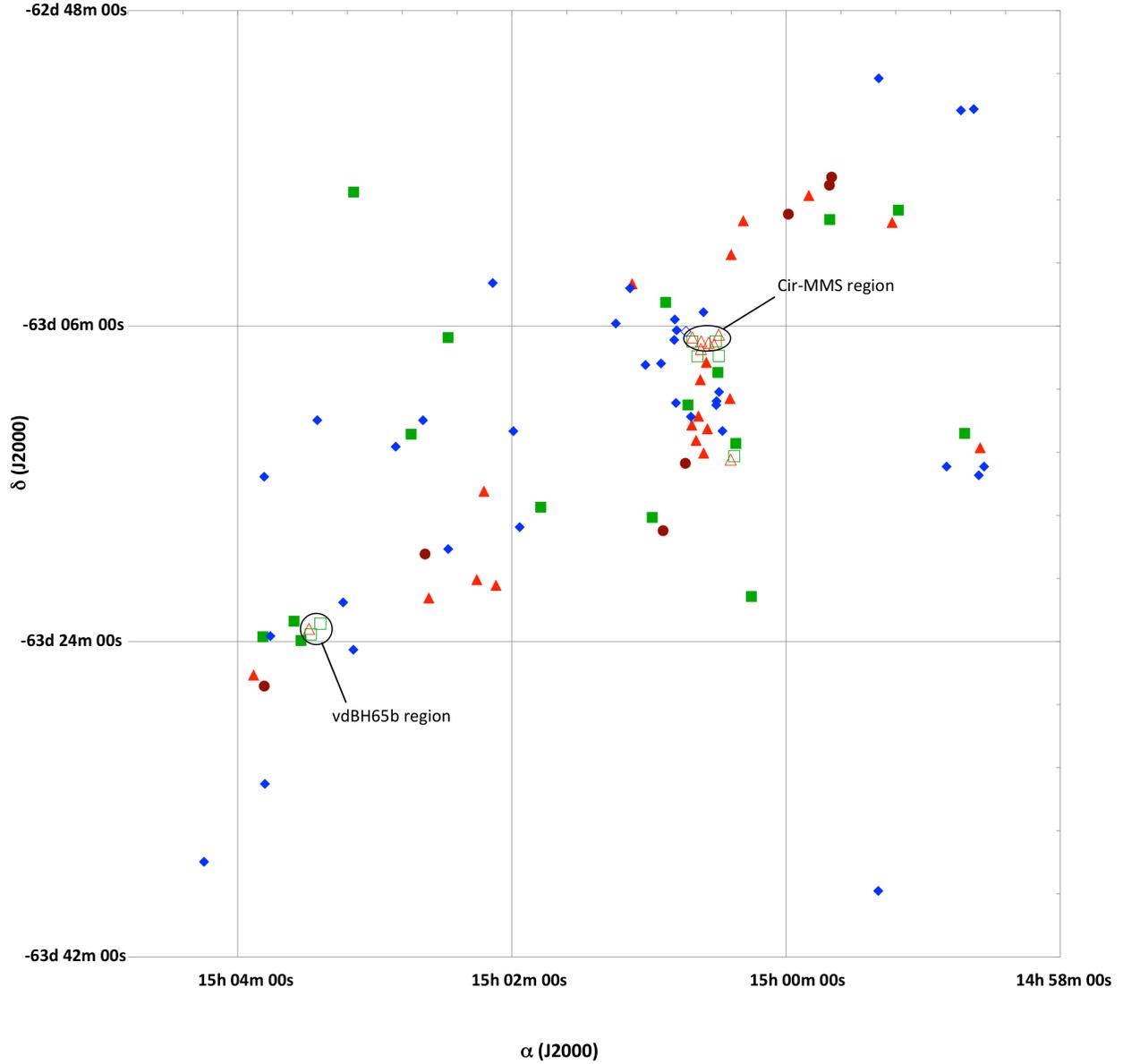}}
\caption{A map of the Class I, flat spectrum, and Class II objects.  \emph{Blue diamonds}- Class II objects; \emph{green squares}- flat spectrum objects; Red triangles- Class I objects.  \emph{Dark red circles} - 'very red' objects (see \S \ref{sec-veryred}).  Open symbols represent crowded objects with uncertain photometry.}
\label{fig-map}
\end{figure}

The distribution of YSOs of different classes can be found in Figure \ref{fig-map}.  Class I and flat spectrum objects, as well as the very red objects, appear to be concentrated in a filament beginning roughly NNW of the Cir-MMS group, continuing to the SSE, with the Cir-MMS region at its center.  This line turns towards the east in the region between the Cir-MMS and vdBH65b aggregates, following a NW to SE line, continuing SE of vdBH65a.  Class II objects appear to be more scattered, with many lying to the east of Cir-MMS and to the north of vdBH65b.

\section{Discussion} \label{sec-disc}
\subsection{Dense Cores and Large Scale Outflows}
Observations by B99 find several large scale CO outflows powered by YSOs in the region.  One of the largest and most energetic of the outflows, designated `Flow A' by B99, originates from the Cir-MMS aggregate (Figure \ref{fig-3color}), slightly west of IRAS 14564-6254.  The nearest bright source to the IRAS object is WISEPC J150033.91-630656.6, though several bright sources lie in this compact aggregate.  This group of sources is also co-located with the Cir-MMS core detected at 1300 $\mu$m (R96).  The WISE observations of the core are consistent with the hypothesis of B99 and R96, that the heating and driving of the outflow are a result of several sources, rather than a single YSO.  This group of YSOs represents the densest collection of luminous mid-infrared sources in the region, with a combined flux density for the region near 10 Jy in Band 4.  The objects in the aggregate are predominantly Class I or flat spectrum objects.   

Flows `B' and `C', originate from IRAS 14563-6301 (B99).  At the location of this source, we find two bright mid-infrared sources close to one another.  They are sources WISEPC J150022.72-631325.4 and J150024.26-631337.4, which are separated by about 14$\arcsec$ (9800 AU at 700 pc). J150024.26-631337.4 is closest to the position of the IRAS source, about 5$\arcsec$ away.  The two WISE sources have 22 $\mu$m fluxes of over 2 Jy and nearly 1 Jy, respectively. IRAS 14563-6301 has a flux of 4.1 Jy at 25 $\mu$m; the discrepancy with WISE photometry can be explained by the larger beam and bandpass of IRAS.  B99 hypothesize that the source of the B and C outflows may be a binary, and this is consistent with our finding of two closely separated mid-infrared sources at the origin.  Several less energetic flows, designated D through H have been identified by B99.  For flows that originate from an uncrowded region, we note the source in Table \ref{tab-sources}. 

\subsection{H$\alpha$ Emitting Stars and Herbig-Haro Objects}
Among the earliest observations of the region were those of \citet{vdbh} which uncovered two stars embedded in nebulosity at optical wavelengths.  Designated 65a and 65b, the latter is associated with an aggregate of YSOs described above, lying near source J150323.80-632258.8 in this study.  Source 65a corresponds to this study's J150058.58-631654.9.  Several candidate YSOs in the Western Circinus cloud have been imaged at 10 $\mu$m by \citet{mottram07}.  One of these sources is vdBH65a, and the other four sources are part of the aggregate near the Cir-MMS core.

A number of other H$\alpha$ emission line stars were uncovered by \citet{mo94}.  Fourteen of the objects they identify are in the region covered by this study, and all have significant mid-infrared excess.  The H$\alpha$ emitting stars are distributed throughout the region, but half of the objects are found near the dense Cir-MMS and vdBH65b aggregates.  A few of these stars are also the source of the large scale CO outflows discussed in the previous section.  Mid-infrared sources that correspond to H$\alpha$ emitting stars are noted in Table \ref{tab-sources}.  

The region also contains several Herbig-Haro objects, specifically HH 76, 77, and 139-143 \citep{rbw08,re94,rg88}.  Several of these sources lie in the vicinity of vdBH65b, as described in \citet{rbw08}.  However, strong mid-infrared tracers of outflows such as FeII lines at 5.3, 18, 26 and 35 $\mu$m and the H$_{2}$ rotational line at 17 $\mu$m \citep{bally07}, do not lie within the four WISE bandpasses.  The WISE images do not show any clear correlation between detected sources and the known HH objects.  

\subsection{Very Red Objects} \label{sec-veryred}
Seven objects are detected as relatively bright sources in Band 4, but have no matching 2MASS sources, and only upper limit measurements in Band 3 (for all but one object, J145959.01-625936.7).  These objects appear as the reddest sources in the 3-color image (Figure \ref{fig-3color}).  Those with Band 1 photometry have [3.4]-[22] $> 8.7$, and all have [4.6]-[22] $> 5.5$.  These sources are noted in Table \ref{tab-sources}, along with their estimated [4.6]-[22] colors.  The list includes one object, J145941.04-625757.6, which was not extracted by the automated pipeline, but was identified by inspection of the image.   These sources are classified as Class I objects based upon $\alpha_{IR} > 0.3$; however, at least one of them (J150238.02-631900.3) may be a Class 0 source, based upon its association with submillimeter emission. 

WISEPC J150238.02-631900.3 corresponds to the location of a dense core described in B99 and may give us insight into the physical nature of these objects.  B99 notes a ``prominent and visually very opaque core which does not harbor a known IR source or [$^{12}$CO outflow]" which they hypothesize to be a pre-collapse core.  The region appears as a dark `knot' in the nebulosity in the WISE image at 12 $\mu$m, but at 22 $\mu$m a source with a flux of about 30 mJy is detected.  The WISE data suggest that this source is possibly in the early stages of evolution, or it could be very deeply embedded, based on its extremely red color.  The reddest object with [4.6]-[22] = 8.5, is J150348.39-632632.3, which is located south of the vdBH65b group.  These objects are shown in the bottom center and right panels of Figure \ref{fig-3color}.  The observational signature of six of the objects is very similar to that of J150238.02-631900.3 (i.e., bright 22 $\mu$m detection, with the 12 $\mu$m image showing a dark knot), which leads us to believe that they may also be physically similar.  J145959.01-625936.7 has a faint 12 $\mu$m detection co-located with the 22 $\mu$m source, and is also located in a dark filamentary structure.  

\subsection{Infrared Luminosities and Comparisons to Nearby Regions}
We calculate the infrared luminosity (L$_{IR}$) for each YSO (48 Class I, flat spectrum, and Class II objects), integrated over the wavelength range 1 to 26 $\mu$m, using the WISE and 2MASS SEDs, and assuming a distance of 700 pc.  Values of L$_{IR}$ are shown in Table \ref{tab-sources} and span the range 0.08 to 30 L$_{\sun}$. A distribution of infrared luminosities is shown in the bottom panel of Figure \ref{fig-cmd}.  The lowest infrared luminosity objects probed by this study are 1 to 2 orders of magnitude more luminous than the lowest-luminosity Spitzer sources in recent studies of nearby regions, such as Serpens, Taurus, and the Cepheus Flare \citep{harvey07,rebull,kirk}. Furthermore, the infrared luminosity distribution of WISE detected sources appears to peak at log(L$_{IR}$/L$_{\sun}$) between -0.5 and 0.0, about an order of magnitude more luminous than those studies.  Thus, the typical YSO observed in this study is significantly more luminous than those observed in the nearby regions.  These effects are due to lower WISE sensitivity compared to Spitzer, and greater distance to the Circinus region.  On the bright end of the luminosity distribution, we find similar numbers of infrared-luminous objects, with the number of YSO falling off sharply where log(L$_{IR}$/L$_{\sun}$) $>$ 0, with no sources more luminous than log(L$_{IR}$/L$_{\sun}$) = 1.5.

The spatial distribution and clustering of YSOs can also be compared qualitatively to Serpens.  The physical size of Serpens is similar to the Western Circinus region.  \citet{harvey07} find that Class II and III objects dominate the population outside of the core, while Class I and flat spectrum objects are common in the core.  We find this also to be the case in the Circinus region, where Class I objects dominate the Cir-MMS and vdbH65b aggregates, and are more confined spatially to dark nebular filaments.  Class II objects are more evenly distributed throughout the region.

\section{Summary}
We have presented WISE observations of the Western Circinus molecular cloud core, covering over 1.1 deg$^{2}$ in the four WISE bands.  These observations have uncovered a population of YSOs with the following characteristics:

- The population contains Class I, II, III and flat spectrum objects. 

- The dense aggregate in the northern region of the cluster, referred to here as the Cir-MMS aggregate, corresponds to a previously discovered dense, cold core at 1300 $\mu$m (R96).  This aggregate also drives a strong CO outflow discovered by B99.

- We identify several mid-infrared sources which correspond to IRAS and H$\alpha$ sources. Some IRAS sources have been resolved into multiple member groups.

- We identify several very red objects with bright detections in Band 4 (22 $\mu$m).  One of these sources corresponds to a dense core of material identified by B99, and could be a Class 0 object.

- The objects observed by WISE in the Circinus region are typically more luminous than those characterized by Spitzer in nearby regions such as Serpens and Taurus.  Clustering of YSO classes is similar to Serpens, where Class I objects are found preferentially in dense cores.

\section{Acknowledgements}
This publication makes use of data products from the Wide-field Infrared Survey Explorer, a joint project of the University of California, Los Angeles, and the Jet Propulsion Laboratory/California Institute of Technology, funded by the National Aeronautics and Space Administration.  W.M.L. acknowledges support from WISE and the WISE Science Data Center.  This research made use of the SIMBAD database, operated at CDS, Strasbourg, France.

\clearpage
\begin{deluxetable}{cccccccccccc}
\tablecaption{WISE Sources in the Western Circinus Cloud}
\tablewidth{0pt}
\rotate
\tabletypesize{\tiny}
\tablehead{
\colhead{ID (WISEPC)$^{1}$} &
\colhead{J (mJy)$^{2}$} &
\colhead{H} &
\colhead{K$_{s}$} &
\colhead{3.4 $\mu$m} &
\colhead{4.6 $\mu$m} &
\colhead{12 $\mu$m} &
\colhead{22 $\mu$m} &
\colhead{$\alpha_{IR}$} &
\colhead{Class} &
\colhead{log(L$_{IR}$/L$_{\sun}$)$^{3}$} &
\colhead{Notes$^{4}$}
}
\startdata
J145716.77-630302.0 & $309 \pm 6.3$ & $660 \pm 17$ & $848 \pm 21$ & $928 \pm 34$ & $646 \pm 13$ & $271 \pm 4.5$ & $126 \pm 2.9$ & -1.82 & III or MS & - & \\
J145726.27-632202.5 & $496 \pm 9.7$ & $807 \pm 36$ & $738 \pm 18$ & $404 \pm 12$ & $192 \pm 3.8$ & $42.6 \pm 0.8$ & $11.1 \pm 0.8$ & -2.8 & III or MS & - & \\
J145726.80-633335.6 & $697 \pm 17$ & $1070 \pm 40$ & $906 \pm 18$ & $465 \pm 13$ & $231 \pm 4.3$ & $48.9 \pm 0.9$ & $15.5 \pm 0.9$ & -2.75 & III or MS & - & \\
J145733.81-630252.3 & $260 \pm 5.8$ & $479 \pm 12$ & $475 \pm 9.3$ & $268 \pm 4.7$ & $125 \pm 2$ & $29.5 \pm 0.6$ & $9.1 \pm 0.7$ & -2.7 & III or MS & - & \\
J145739.30-631851.7 & $2040 \pm 46$ & $3860 \pm 120$ & $3920 \pm 88$ & $1870 \pm 92$ & $1250 \pm 35$ & $252 \pm 4.2$ & $80.5 \pm 1.8$ & -2.67 & III or MS & - & \\
J145740.79-633533.3 & $195 \pm 4.4$ & $373 \pm 16$ & $395 \pm 12$ & $248 \pm 2.8$ & $129 \pm 1.7$ & $37.5 \pm 0.6$ & $15.1 \pm 0.7$ & -2.4 & III or MS & - & \\
J145753.68-625737.3 & $2200 \pm 60$ & $3670 \pm 120$ & $3370 \pm 63$ & $1730 \pm 83$ & $963 \pm 22$ & $192 \pm 3.4$ & $57.6 \pm 1.8$ & -2.75 & III or MS & - & \\
J145753.92-631233.9 & $73.5 \pm 1.7$ & $117 \pm 2.4$ & $107 \pm 1.9$ & $58.4 \pm 1.2$ & $30 \pm 0.5$ & $9.2 \pm 0.2$ & $14.5 \pm 0.8$ & -1.86 & III or MS & - & \\
J145757.66-633055.5 & $952 \pm 23$ & $1720 \pm 58$ & $1780 \pm 38$ & $1010 \pm 38$ & $491 \pm 9.6$ & $105 \pm 1.8$ & $30.1 \pm 1.1$ & -2.75 & III or MS & - & \\
J145758.89-631715.6 & $1830 \pm 36$ & $1380 \pm 44$ & $977 \pm 15$ & $450 \pm 12$ & $246 \pm 5$ & $45.5 \pm 0.8$ & $11.1 \pm 0.8$ & -2.92 & III or MS & - & \\
J145805.27-631709.0 & $234 \pm 5$ & $493 \pm 20$ & $508 \pm 11$ & $272 \pm 5.8$ & $137 \pm 2.5$ & $31.3 \pm 0.6$ & $9.3 \pm 0.8$ & -2.72 & III or MS & - & \\
J145810.30-625201.6 & $233 \pm 4.7$ & $623 \pm 23$ & $738 \pm 12$ & $463 \pm 9$ & $306 \pm 5.4$ & $95 \pm 1.5$ & $44.5 \pm 1.2$ & -2.21 & III or MS & - & \\
J145812.83-631236.3 & $669 \pm 14$ & $1080 \pm 23$ & $926 \pm 15$ & $482 \pm 12$ & $260 \pm 4.6$ & $52.7 \pm 0.9$ & $14.4 \pm 0.8$ & -2.79 & III or MS & - & \\
J145815.57-630219.2 & $192 \pm 4.1$ & $381 \pm 11$ & $383 \pm 8.2$ & $214 \pm 4.7$ & $110 \pm 2.1$ & $24.9 \pm 0.4$ & $9.5 \pm 0.8$ & -2.59 & III or MS & - & \\
J145827.41-625203.0 & $1180 \pm 26$ & $2170 \pm 90$ & $2250 \pm 44$ & $898 \pm 19$ & $590 \pm 9.3$ & $138 \pm 2.3$ & $41.1 \pm 1.2$ & -2.72 & III or MS & - & \\
J145831.10-625449.8 & $671 \pm 14$ & $2400 \pm 110$ & $3500 \pm 95$ & $2440 \pm 130$ & $1990 \pm 60$ & $468 \pm 7.3$ & $235 \pm 5.2$ & -2.16 & III or MS & - & \\
J145832.83-630408.8 & $2490 \pm 56$ & $5050 \pm 210$ & $6010 \pm 95$ & $3160 \pm 150$ & $2230 \pm 69$ & $1030 \pm 15$ & $741 \pm 13$ & -1.9 & III or MS & - & IRAS 14544-6252\\
J145833.36-631401.0 & $3.9 \pm 0.1$ & $9 \pm 0.2$ & $10.2 \pm 0.2$ & $11.2 \pm 0.3$ & $11.6 \pm 0.2$ & $17.3 \pm 0.3$ & $45.3 \pm 0.7$ & -0.36 & II & -0.38 & \\
J145835.09-631256.5 & - & $0.9 \pm 0.03$ & $3.9 \pm 0.2$ & $9.5 \pm 0.3$ & $24.9 \pm 0.5$ & $82.2 \pm 1.3$ & $283 \pm 5$ & 0.84 & I & - & \\
J145835.68-631430.8 & $3.3 \pm 0.1$ & $5.3 \pm 0.2$ & $4.8 \pm 0.2$ & $4.8 \pm 0.2$ & $4.3 \pm 0.1$ & $7.1 \pm 0.2$ & $17.6 \pm 0.8$ & -0.44 & II & -0.71 & \\
J145837.86-625337.2 & $4.8 \pm 0.2$ & $9.1 \pm 0.2$ & $11.5 \pm 0.3$ & $6.5 \pm 0.2$ & $7.5 \pm 0.1$ & $13.5 \pm 0.3$ & $15.9 \pm 0.9$ & -0.86 & II & -0.47 & \\
J145841.36-631048.9 & $182 \pm 4$ & $421 \pm 33$ & $503 \pm 11$ & $279 \pm 4.1$ & $148 \pm 2.2$ & $38.5 \pm 0.8$ & $15.1 \pm 0.9$ & -2.51 & III or MS & - & \\
J145841.86-631207.5 & - & - & - & $4 \pm 0.1$ & $5.6 \pm 0.1$ & $10.2 \pm 0.2$ & $35.1 \pm 0.6$ & 0.15 & Flat & - & \\
J145843.48-625341.5 & $87.8 \pm 1.8$ & $132 \pm 2.7$ & $148 \pm 2.6$ & $155 \pm 3$ & $160 \pm 2.5$ & $135 \pm 1.8$ & $145 \pm 2.3$ & -1.01 & II & 0.71 & MO94-1\\
J145848.44-630657.1 & $133 \pm 2.8$ & $263 \pm 5.4$ & $240 \pm 5.4$ & $136 \pm 1.1$ & $68.3 \pm 0.4$ & $16.5 \pm 0.2$ & $7.4 \pm 0.5$ & -2.49 & III or MS & - & \\
J145849.77-631401.0 & $98.3 \pm 2.2$ & $294 \pm 6.8$ & $600 \pm 19$ & $633 \pm 21$ & $601 \pm 13$ & $510 \pm 8.1$ & $369 \pm 6.5$ & -1.21 & II & 1.23 & IRAS 14547-6302\\
J145856.01-630822.4 & $291 \pm 6$ & $626 \pm 19$ & $626 \pm 15$ & $376 \pm 9.1$ & $205 \pm 3.8$ & $78.2 \pm 1.4$ & $45.9 \pm 1.3$ & -2.12 & III or MS & - & \\
J145901.82-633252.6 & $143 \pm 3.4$ & $267 \pm 14$ & $285 \pm 11$ & $150 \pm 3.5$ & $131 \pm 2.6$ & $61.3 \pm 1$ & $28.2 \pm 1.1$ & -1.99 & III or MS & - & \\
J145903.59-632507.4 & $16300 \pm 670$ & $30500 \pm 8000$ & $29600 \pm 8900$ & $16800 \pm 790$ & $15900 \pm 220$ & $4050 \pm 83$ & $1330 \pm 22$ & -2.33 & III or MS & - & IRAS 14549-6313\\
J145910.87-625923.3 & $1.2 \pm 0.1$ & $2.2 \pm 0.2$ & $2.5 \pm 0.2$ & $2.7 \pm 0.04$ & $3 \pm 0.1$ & $8.6 \pm 0.3$ & $17.5 \pm 0.8$ & -0.16 & Flat & -0.89 & \\
J145913.60-630005.5 & $9.4 \pm 0.2$ & $14.2 \pm 0.5$ & $15.1 \pm 0.3$ & $15.7 \pm 0.4$ & $25.1 \pm 0.4$ & $212 \pm 0.8$ & $675 \pm 1.3$ & 0.64 & I & 0.31 & IRAS 14551-6248; MO94-2\\
J145919.46-625151.8 & $2.1 \pm 0.1$ & $3.9 \pm 0.2$ & $3.6 \pm 0.2$ & $2.1 \pm 0.04$ & $1.1 \pm 0.04$ & $3.3 \pm 0.1$ & $11.6 \pm 0.9$ & -0.5 & II & -0.95 & \\
J145919.65-633813.7 & $154 \pm 3.7$ & $456 \pm 16$ & $794 \pm 17$ & $816 \pm 27$ & $1120 \pm 28$ & $1660 \pm 29$ & $1090 \pm 17$ & -0.86 & II & 1.47 & IRAS 14551-6326\\
J145922.50-632710.9 & $351 \pm 7.8$ & $726 \pm 27$ & $682 \pm 13$ & $388 \pm 9$ & $204 \pm 3.8$ & $44 \pm 0.8$ & $11 \pm 0.9$ & -2.77 & III or MS & - & \\
J145923.06-630322.6 & $124 \pm 2.7$ & $330 \pm 7.4$ & $476 \pm 11$ & $340 \pm 7.3$ & $301 \pm 5$ & $141 \pm 2.4$ & $71.2 \pm 1.8$ & -1.82 & III or MS & - & \\
J145924.41-630832.5 & $55.3 \pm 1.3$ & $151 \pm 3.4$ & $193 \pm 3.7$ & $140 \pm 2.8$ & $91.1 \pm 1.6$ & $34.9 \pm 0.6$ & $23.8 \pm 0.9$ & -1.9 & III or MS & - & \\
J145925.01-633906.9 & $1290 \pm 24$ & $2480 \pm 88$ & $2690 \pm 40$ & $1380 \pm 5.1$ & $837 \pm 3.8$ & $393 \pm 3.2$ & $272 \pm 1.7$ & -1.98 & III or MS & - & \\
J145925.08-630018.9 & $38.3 \pm 0.9$ & $77.9 \pm 1.7$ & $77.5 \pm 1.7$ & $45.3 \pm 1$ & $27.7 \pm 0.5$ & $7.5 \pm 0.2$ & $17.9 \pm 0.9$ & -1.63 & III or MS & - & \\
J145925.39-625121.6 & $240 \pm 5.4$ & $520 \pm 17$ & $579 \pm 11$ & $339 \pm 8.2$ & $173 \pm 2.9$ & $50.5 \pm 0.9$ & $24.6 \pm 1.1$ & -2.36 & III or MS & - & \\
J145925.58-633714.1 & $415 \pm 12$ & $759 \pm 26$ & $760 \pm 15$ & $416 \pm 11$ & $197 \pm 3.8$ & $47.9 \pm 0.8$ & $12.2 \pm 0.9$ & -2.78 & III or MS & - & \\
J145926.13-633608.4 & $203 \pm 4.6$ & $361 \pm 10$ & $394 \pm 11$ & $266 \pm 4.7$ & $176 \pm 2.8$ & $77 \pm 1.1$ & $39.3 \pm 0.8$ & -1.99 & III or MS & - & \\
J145936.09-630828.9 & $128 \pm 2.9$ & $449 \pm 18$ & $613 \pm 13$ & $373 \pm 5.6$ & $209 \pm 3.1$ & $47.7 \pm 0.8$ & $12.5 \pm 0.7$ & -2.67 & III or MS & - & \\
J145936.85-625439.9 & $426 \pm 9.1$ & $834 \pm 30$ & $821 \pm 13$ & $473 \pm 13$ & $220 \pm 4.1$ & $52 \pm 1$ & $16.5 \pm 1$ & -2.68 & III or MS & - & \\
J145940.08-625730.0 & $-  \pm -$ & - & - & $0.1 \pm 0.1$ & $2 \pm 0.1$ & 0.25u & $16.8 \pm 0.9$ & 1.71 & I & - & very red; [4.6]-[22]=5.57\\
J145940.73-632055.8 & $1350 \pm 34$ & $2730 \pm 74$ & $3060 \pm 57$ & $1590 \pm 79$ & $1410 \pm 38$ & $399 \pm 6.7$ & $191 \pm 4.5$ & -2.19 & III or MS & - & IRAS 14555-6308\\
J145940.94-625955.0 & $0.6 \pm 0.1$ & $1.2 \pm 0.1$ & $1.8 \pm 0.2$ & $2 \pm 0.04$ & $1.5 \pm 0.1$ & $6 \pm 0.2$ & $10.1 \pm 0.8$ & -0.26 & Flat & -1.09 & \\
J145941.04-625757.6 & - & - & - & - & - & - & - & - & N/A & - & very red; no photometry\\
J145944.57-631131.8 & $662 \pm 16$ & $1700 \pm 64$ & $2170 \pm 63$ & $1240 \pm 52$ & $935 \pm 23$ & $401 \pm 6.7$ & $275 \pm 4.9$ & -1.89 & III or MS & - & IRAS 14556-6259\\
J145945.97-631434.4 & $1320 \pm 25$ & $3140 \pm 91$ & $3830 \pm 85$ & $2400 \pm 120$ & $1700 \pm 52$ & $749 \pm 13$ & $643 \pm 11$ & -1.77 & III or MS & - & IRAS 14556-6302\\
J145949.82-632242.8 & $387 \pm 9.8$ & $791 \pm 28$ & $787 \pm 17$ & $423 \pm 7.8$ & $208 \pm 3.1$ & $47.4 \pm 0.8$ & $13.5 \pm 0.7$ & -2.75 & III or MS & - & \\
J145950.07-625833.1 & - & - & - & $0.6 \pm 0.2$ & $2.4 \pm 0.2$ & $11.1 \pm 0.6$ & $84.9 \pm 4$ & 1.63 & I & - & \\
J145951.15-625851.3 & $1580 \pm 34$ & $1680 \pm 42$ & $1230 \pm 36$ & $605 \pm 16$ & $318 \pm 5.9$ & $61.9 \pm 1$ & $23.6 \pm 1$ & -2.7 & III or MS & - & \\
J145955.31-630538.5 & $1500 \pm 25$ & $1630 \pm 40$ & $1150 \pm 28$ & $570 \pm 18$ & $293 \pm 6$ & $57.4 \pm 1.1$ & $12.8 \pm 0.8$ & -2.93 & III or MS & - & \\
J145959.01-625936.7 & - & - & - & $0.6 \pm 0.002$ & $2.8 \pm 0.1$ & 0.32u & $48.2 \pm 1.3$ & 1.33 & I & - & very red; [4.6]-[22]=6.38\\
J145959.30-631309.7 & $25.9 \pm 0.6$ & $153 \pm 3.1$ & $284 \pm 5.6$ & $207 \pm 0.6$ & $157 \pm 0.4$ & $46 \pm 0.3$ & $21.1 \pm 0.5$ & -2.12 & III or MS & - & \\
J150000.30-632011.9 & $103 \pm 2.2$ & $213 \pm 4.3$ & $223 \pm 4.8$ & $136 \pm 2.7$ & $82.1 \pm 1.3$ & $38.8 \pm 0.7$ & $20.7 \pm 1.1$ & -2.02 & III or MS & - & \\
J150000.78-631853.6 & $130 \pm 2.8$ & $377 \pm 16$ & $475 \pm 7.5$ & $278 \pm 6$ & $151 \pm 2.8$ & $33.8 \pm 0.6$ & $11.1 \pm 0.8$ & -2.61 & III or MS & - & \\
J150002.74-632201.2 & $263 \pm 5.9$ & $612 \pm 23$ & $736 \pm 14$ & $407 \pm 11$ & $225 \pm 4.4$ & $124 \pm 2$ & $87.9 \pm 2.4$ & -1.91 & III or MS & - & \\
J150003.89-630004.9 & $8.6 \pm 0.2$ & $130 \pm 2.5$ & $441 \pm 8.6$ & $391 \pm 10$ & $336 \pm 6.2$ & $89.9 \pm 1.6$ & $39.1 \pm 1.3$ & -2.04 & III or MS & - & \\
J150004.33-625533.0 & $103 \pm 2$ & $230 \pm 4.5$ & $258 \pm 5$ & $156 \pm 2.6$ & $84.8 \pm 1.3$ & $24.5 \pm 0.5$ & $10.6 \pm 0.8$ & -2.37 & III or MS & - & \\
J150009.75-633242.2 & $554 \pm 16$ & $1000 \pm 44$ & $982 \pm 19$ & $525 \pm 11$ & $259 \pm 4.3$ & $57.4 \pm 1.1$ & $15.5 \pm 0.9$ & -2.78 & III or MS & - & \\
J150015.19-632125.9 & - & - & $0.7 \pm 0.2$ & $0.8 \pm 0.03$ & $0.7 \pm 0.04$ & $4.4 \pm 0.2$ & $10.4 \pm 0.9$ & 0.16 & Flat & - & \\
J150017.88-633544.9 & $865 \pm 16$ & $1610 \pm 70$ & $2060 \pm 32$ & $965 \pm 35$ & $871 \pm 20$ & $359 \pm 5.7$ & $167 \pm 3.4$ & -2.08 & III or MS & - & \\
J150018.73-625959.7 & - & - & - & $0.8 \pm 0.03$ & $7.4 \pm 0.1$ & $32.5 \pm 0.5$ & $1370 \pm 22$ & 2.95 & I & - & IRAS 14562-6248\\
J150021.82-630220.3 & $2.5 \pm 0.1$ & $42.5 \pm 0.9$ & $150 \pm 3.2$ & $123 \pm 2.3$ & $104 \pm 1.7$ & $25.8 \pm 0.5$ & $18.4 \pm 1$ & -1.9 & III or MS & -& \\
J150022.02-631241.9 & $0.6 \pm 0.1$ & $6 \pm 0.1$ & $29.5 \pm 0.6$ & $87.4 \pm 2$ & $167 \pm 3.2$ & $202 \pm 3.8$ & $406 \pm 7.2$ & 0.13 & Flat & 0.46 & \\
J150022.72-631325.4 & $63.9 \pm -$ & $185.7 \pm 5.7$ & $345 \pm 6.4$ & $\sim$221 & $\sim$324 & $\sim$912 & $\sim$2425 & $\sim$-0.2 & Flat? & - & C\\
J150024.04-630155.5 & - & - & - & $46.9 \pm 1.1$ & $311 \pm 5.8$ & $428 \pm 6.7$ & $1200 \pm 17$ & 0.72 & I & - & IRAS 14563-6250\\
J150024.26-631337.4 & - & - & - & $\sim$18 & $\sim$31 & $\sim$58 & $\sim$969 & $\sim$1.1 & I? & - & C\\
J150024.55-631008.2 & $0.6 \pm 0.03$ & $5.3 \pm 0.1$ & $25.2 \pm 0.6$ & $55.6 \pm 1.3$ & $113 \pm 2.3$ & $241 \pm 4.1$ & $609 \pm 11$ & 0.37 & I & 0.44 & C$\arcmin$\\
J150025.45-632945.0 & $221 \pm 4.8$ & $399 \pm 16$ & $375 \pm 10$ & $201 \pm 3.5$ & $93.4 \pm 1.5$ & $22.3 \pm 0.4$ & $9.9 \pm 0.8$ & -2.56 & III or MS & - & \\
J150027.85-631159.1 & $5 \pm 0.1$ & $9.7 \pm 0.2$ & $11.6 \pm 0.3$ & $9 \pm 0.2$ & $8.6 \pm 0.2$ & $10.5 \pm 0.3$ & $16.2 \pm 1.3$ & -0.86 & II & -0.45 & \\
J150027.96-633620.5 & $961 \pm 19$ & $1860 \pm 81$ & $2050 \pm 32$ & $1110 \pm 47$ & $641 \pm 14$ & $185 \pm 3.1$ & $82.9 \pm 2$ & -2.38 & III or MS & - & \\
J150029.34-630945.5 & $23.7 \pm 0.8$ & $37.3 \pm 1.7$ & $51.9 \pm 1.3$ & $49.1 \pm 1.1$ & $45.7 \pm 0.8$ & $60.5 \pm 0.9$ & $98.9 \pm 1.7$ & -0.72 & II & 0.23 & MO94-4\\
J150029.44-630742.3 & 0.12u & $1.3 \pm 0.1$ & $21.7 \pm 0.5$ & $\sim$134 & $\sim$545 & $\sim$913 & $\sim$3087 & $\sim$0.1 & Flat? & - & C; Cir-MMS\\
J150029.46-630629.4 & $-  \pm -$ & - & - & $\sim$4 & $\sim$7 & $\sim$19 & $\sim$100 & $\sim$0.7 & I? & - & C; Cir-MMS\\
J150029.84-630838.5 & $-  \pm -$ & - & $1.6 \pm 0.1$ & $1.9 \pm 0.1$ & $6.1 \pm 0.1$ & $7.8 \pm 0.2$ & $27.2 \pm 1.6$ & 0.22 & Flat & - & \\
J150030.26-631418.4 & $491 \pm 10$ & $1770 \pm 73$ & $2960 \pm 58$ & $1690 \pm 78$ & $1510 \pm 41$ & $496 \pm 8.7$ & $266 \pm 6.2$ & -2.04 & III or MS & - & \\
J150030.42-631017.0 & $6.7 \pm 0.1$ & $29.1 \pm 0.6$ & $43.8 \pm 0.8$ & $24.5 \pm 0.2$ & $17.4 \pm 0.1$ & $7 \pm 0.2$ & $19.3 \pm 0.7$ & -1.35 & II & -0.05 & \\
J150030.57-631030.3 & $1.2 \pm 0.01$ & $7.8 \pm 0.2$ & $17.8 \pm 0.3$ & $19.3 \pm 0.4$ & $25.2 \pm 0.5$ & $20.4 \pm 0.4$ & $41.6 \pm 1$ & -0.63 & II & -0.24 & \\
J150030.78-630652.2 & $41.6 \pm 0.9$ & $65.8 \pm 1.5$ & $72.4 \pm 1.5$ & $\sim$76 & $\sim$89 & $\sim$188 & $\sim$450 & $\sim$-0.2 & Flat? & - & C; Cir-MMS\\
J150031.50-630703.3 & - & - & - & $\sim$21 & $\sim$55 & $\sim$175 & $\sim$895 & $\sim$1.0 & I? & - & C; Cir-MMS\\
J150033.91-630656.6 & 0.51u & $0.9 \pm 0.1$ & $3.7 \pm 0.2$ & $\sim$37 & $\sim$221 & $\sim$1174 & $\sim$5044 & $\sim$1.1 & I? & - & C; Cir-MMS\\
J150034.43-631151.9 & - & - & - & $1.8 \pm 0.03$ & $6.9 \pm 0.1$ & $7.8 \pm 0.3$ & $216 \pm 5.2$ & 1.54 & I & - & \\
J150034.91-630804.4 & - & - & - & $0.5 \pm 0.1$ & $3.2 \pm 0.1$ & $4.7 \pm 0.3$ & $37.9 \pm 1.7$ & 1.3 & I & - & \\
J150036.06-631315.0 & $0.4 \pm 0.1$ & $1.4 \pm 0.1$ & $4.6 \pm 0.2$ & $11.1 \pm 0.2$ & $35 \pm 0.6$ & $92.5 \pm 0.9$ & $319 \pm 2$ & 0.82 & I & -0.01 & \\
J150036.15-630512.4 & $22.5 \pm 0.6$ & $41.6 \pm 1.2$ & $45.3 \pm 1$ & $27.4 \pm 0.6$ & $21.1 \pm 0.4$ & $14.5 \pm 0.4$ & $29.3 \pm 1.3$ & -1.19 & II & 0.07 & \\
J150037.15-630652.2 & $0.4 \pm 0.1$ & $5.5 \pm 0.1$ & $29.4 \pm 0.6$ & $\sim$51 & $\sim$112 & $\sim$182 & $\sim$415 & $\sim$2.1 & I? & - & C; Cir-MMS\\
J150037.28-630718.2 & 0.37u & 0.77u & $3.1 \pm 0.1$ & $\sim$15 & $\sim$85 & $\sim$126 & $\sim$570 & $\sim$2.4 & I? & - & C; Cir-MMS\\
J150037.54-630903.6 & - & $0.9 \pm 0.05$ & $4 \pm 0.2$ & $5.9 \pm 0.1$ & $14 \pm 0.3$ & $28.9 \pm 0.6$ & $177 \pm 4$ & 0.63 & I & - & \\
J150038.34-631108.2 & - & - & - & $2.2 \pm 0.1$ & $13.8 \pm 0.2$ & $33.9 \pm 0.6$ & $342 \pm 6.3$ & 1.68 & I & - & \\
J150038.74-630743.3 & - & - & $2.9 \pm 0.1$ & $\sim$12 & $\sim$37 & $\sim$8 & $\sim$30 & $\sim$0.0 & Flat & - & C; Cir-MMS\\
J150039.38-631231.5 & - & $0.7 \pm 0.1$ & $8 \pm 0.3$ & $30.5 \pm 0.7$ & $82 \pm 1.6$ & $139.3 \pm 2.6$ & $341 \pm 7.9$ & 0.61 & I & - & \\
J150040.62-631522.6 & $51.4 \pm 1.1$ & $129 \pm 2.6$ & $160 \pm 2.8$ & $92.9 \pm 2.1$ & $68 \pm 1.2$ & $29.3 \pm 0.5$ & $15.5 \pm 0.9$ & -2 & III or MS & - & \\
J150041.0-630652.9 & 0.18u & $2.3 \pm 0.2$ & $8 \pm 0.4$ & $\sim$16 & $\sim$34 & $\sim$50 & $\sim$109 & $\sim$0.1 & Flat? & - & C; Cir-MMS\\
J150041.05-630638.9 & 0.08u & $0.8 \pm 0.03$ & $1.8 \pm 0.2$ & $\sim$7 & $\sim$87 & $\sim$707 & $\sim$4695 & $\sim$1.2 & I? & - & C; Cir-MMS\\
J150041.29-631138.9 & - & - & - & $1.3 \pm 0.1$ & $13.8 \pm 0.3$ & $30.2 \pm 0.6$ & $225 \pm 3.9$ & 1.73 & I & - & \\
J150041.66-631110.3 & $29.7 \pm 0.8$ & $71.5 \pm 1.5$ & $110 \pm 2.4$ & $140 \pm 2.9$ & $142 \pm 3$ & $199 \pm 3.6$ & $288 \pm 5.9$ & -0.58 & II & 0.63 & \\
J150042.94-631030.1 & $2 \pm 0.1$ & $4.3 \pm 0.1$ & $5.9 \pm 0.1$ & $4.9 \pm 0.1$ & $6.3 \pm 0.1$ & $6.5 \pm 0.2$ & $50.1 \pm 1.4$ & -0.08 & Flat & -0.65 & \\
J150043.60-631731.9 & $216 \pm 4.4$ & $630 \pm 30$ & $968 \pm 14$ & $734 \pm 23$ & $611 \pm 12$ & $244 \pm 4.1$ & $116 \pm 2.9$ & -1.91 & III or MS & - & \\
J150043.83-630618.4 & $3.3 \pm 0.1$ & $9.1 \pm 0.3$ & $12.8 \pm 0.3$ & $\sim$12 & $\sim$12 & $\sim$13 & $\sim$47 & $\sim$-0.4 & II? & - & C; Cir-MMS1\\
J150044.06-630106.9 & $60.3 \pm 1.3$ & $148 \pm 3.1$ & $175 \pm 3.1$ & $102 \pm 2.2$ & $59.1 \pm 1.1$ & $25.7 \pm 0.5$ & $17.3 \pm 0.8$ & -1.99 & III or MS & - & \\
J150044.08-631349.7 & - & - & - & $0.1 \pm 0.05$ & $1.8 \pm 0.04$ & 0.25u & $13.7 \pm 1$ & 1.61 & I & - & very red; [4.6]-[22]=5.50\\
J150044.61-633851.4 & $831 \pm 18$ & $1550 \pm 68$ & $1580 \pm 50$ & $849 \pm 28$ & $429 \pm 8.4$ & $103 \pm 1.8$ & $31 \pm 1.1$ & -2.69 & III or MS & - & \\
J150046.01-625427.2 & $3920 \pm 77$ & $4480 \pm 190$ & $3640 \pm 78$ & $1270 \pm 22$ & $955 \pm 12$ & $203 \pm 2.4$ & $49.8 \pm 1.2$ & -2.84 & III or MS & - & \\
J150047.90-630614.3 & $5.4 \pm 0.2$ & $12.4 \pm 0.4$ & $15.6 \pm 0.4$ & $14.6 \pm 0.4$ & $13.2 \pm 0.3$ & $9 \pm 0.3$ & $16.7 \pm 1.4$ & -0.97 & II & -0.34 & \\
J150048.20-631023.1 & $0.6 \pm 0.1$ & $2.7 \pm 0.1$ & $5.4 \pm 0.1$ & $6.8 \pm 0.2$ & $7.7 \pm 0.2$ & $4.1 \pm 0.2$ & $8.1 \pm 0.7$ & -0.82 & II & -0.77 & \\
J150048.68-630537.3 & $60.3 \pm 1.2$ & $94.8 \pm 1.9$ & $121 \pm 2.5$ & $112 \pm 2.5$ & $106 \pm 2$ & $178 \pm 2.1$ & $226 \pm 2.3$ & -0.73 & II & 0.62 & MO94-7\\
J150048.95-630647.4 & $5.3 \pm 0.2$ & $11.1 \pm 0.2$ & $14.9 \pm 0.3$ & $16 \pm 0.5$ & $16.1 \pm 0.3$ & $20.5 \pm 0.5$ & $55 \pm 2.2$ & -0.44 & II & -0.26 & MO94-8\\
J150052.69-630438.8 & - & - & - & $1.3 \pm 0.1$ & $2.6 \pm 0.05$ & $5.3 \pm 0.2$ & $14.3 \pm 0.8$ & 0.27 & Flat & - & \\
J150053.83-631740.4 & - & - & - & 0.11u & $0.3 \pm 0.02$ & 0.27u & $18.7 \pm 1.1$ & - & N/A & - & very red; [4.6]-[22]=7.87\\
J150054.73-630807.9 & $4.1 \pm 0.1$ & $8.4 \pm 0.2$ & $11.7 \pm 0.3$ & $9.9 \pm 0.2$ & $10.7 \pm 0.2$ & $12.6 \pm 0.3$ & $18.3 \pm 1$ & -0.81 & II & -0.44 & \\
J150057.48-633144.0 & $273 \pm 5.9$ & $500 \pm 23$ & $542 \pm 12$ & $298 \pm 6.7$ & $148 \pm 2.7$ & $36 \pm 0.7$ & $11.9 \pm 0.8$ & -2.64 & III or MS & - & \\
J150058.58-631654.9 & $32.3 \pm 0.9$ & $96.5 \pm 2.3$ & $208 \pm 3.7$ & $277 \pm 7$ & $370 \pm 6.9$ & $1070 \pm 18$ & $2240 \pm 37$ & 0.02 & Flat & 1.11 & IRAS 14568-6304; vdBH65a; MO94-10; Flow F\\
J150100.11-625427.0 & $12500 \pm 370$ & $20000 \pm 3800$ & $18800 \pm 5100$ & $12400 \pm 750$ & $9330 \pm 300$ & $1390 \pm 26$ & $434 \pm 8.1$ & -2.62 & III or MS & - & \\
J150101.54-630813.0 & $6.3 \pm 0.1$ & $12.4 \pm 0.2$ & $16.7 \pm 0.4$ & $14.1 \pm 0.4$ & $13.4 \pm 0.3$ & $9.4 \pm 0.3$ & $12.8 \pm 0.8$ & -1.11 & II & -0.33 & \\
J150104.54-625557.2 & $504 \pm 13$ & $722 \pm 32$ & $621 \pm 10$ & $338 \pm 3.5$ & $172 \pm 1.9$ & $34.2 \pm 0.7$ & $8.9 \pm 0.7$ & -2.82 & III or MS & - & \\
J150107.38-630335.4 & - & $0.7 \pm 0.1$ & $1.4 \pm 0.1$ & $5.1 \pm 0.1$ & $7.9 \pm 0.1$ & $23.2 \pm 0.5$ & $39.4 \pm 1$ & 0.44 & I & - & \\
J150108.29-630350.5 & $13 \pm 0.3$ & $21.3 \pm 0.5$ & $21.5 \pm 0.5$ & $16 \pm 0.4$ & $12.7 \pm 0.3$ & $8.5 \pm 0.2$ & $14 \pm 0.5$ & -1.19 & II & -0.21 & \\
J150112.08-633014.6 & $820 \pm 18$ & $1260 \pm 58$ & $1170 \pm 22$ & $625 \pm 19$ & $300 \pm 5.6$ & $65.2 \pm 1$ & $17.8 \pm 0.9$ & -2.8 & III or MS & - & \\
J150114.50-630551.1 & $12.2 \pm 0.2$ & $18.9 \pm 0.3$ & $21.2 \pm 0.4$ & $19.1 \pm 0.5$ & $15.8 \pm 0.4$ & $9.6 \pm 0.3$ & $28.6 \pm 1.1$ & -0.87 & II & -0.19 & \\
J150115.48-631202.5 & $48 \pm 1$ & $105 \pm 2.1$ & $112 \pm 2$ & $64 \pm 1.5$ & $39.2 \pm 0.9$ & $15.3 \pm 0.3$ & $9.3 \pm 0.8$ & -2.07 & III or MS & - & \\
J150117.08-625646.6 & $94.4 \pm 2.1$ & $192 \pm 3.9$ & $218 \pm 4.9$ & $129 \pm 2.6$ & $67.8 \pm 1.3$ & $40.6 \pm 0.7$ & $31.9 \pm 0.9$ & -1.83 & III or MS & - & \\
J150119.42-630223.3 & $116 \pm 2.6$ & $291 \pm 17$ & $372 \pm 11$ & $241 \pm 5.4$ & $143 \pm 2.8$ & $71.2 \pm 1.3$ & $41.4 \pm 1.2$ & -1.94 & III or MS & - & \\
J150122.83-633850.5 & $1670 \pm 47$ & $2900 \pm 100$ & $2700 \pm 50$ & $1400 \pm 60$ & $719 \pm 17$ & $153 \pm 2.6$ & $41.8 \pm 1.5$ & -2.79 & III or MS & - & \\
J150124.66-631608.7 & $820 \pm 20$ & $1740 \pm 55$ & $1840 \pm 29$ & $1040 \pm 40$ & $561 \pm 12$ & $118 \pm 2$ & $35.7 \pm 1.3$ & -2.69 & III or MS & - & \\
J150126.44-630707.1 & $100 \pm 2.2$ & $218 \pm 4.5$ & $241 \pm 4.7$ & $157 \pm 3.3$ & $96 \pm 1.8$ & $33.8 \pm 0.6$ & $17.1 \pm 0.8$ & -2.14 & III or MS & - & \\
J150130.91-631901.4 & $258 \pm 5.5$ & $561 \pm 27$ & $652 \pm 20$ & $273 \pm 6.1$ & $236 \pm 4.4$ & $92.6 \pm 1.7$ & $43.3 \pm 1.4$ & -2.16 & III or MS & - & \\
J150140.06-630253.7 & $2330 \pm 52$ & $6110 \pm 260$ & $7500 \pm 130$ & $4400 \pm 240$ & $4890 \pm 160$ & $780 \pm 12$ & $280 \pm 4.9$ & -2.41 & III or MS & - & \\
J150144.37-631429.8 & $1490 \pm 38$ & $2680 \pm 60$ & $2590 \pm 51$ & $1300 \pm 50$ & $740 \pm 17$ & $158 \pm 2.7$ & $46.1 \pm 1.5$ & -2.73 & III or MS & - & \\
J150147.36-631620.2 & $1.2 \pm 0.1$ & $2.8 \pm 0.2$ & $5 \pm 0.2$ & $9.5 \pm 0.2$ & $12.5 \pm 0.2$ & $20.5 \pm 0.4$ & $57.2 \pm 1.8$ & 0.05 & Flat & -0.48 & \\
J150148.30-630343.8 & $661 \pm 16$ & $1330 \pm 45$ & $1390 \pm 43$ & $487 \pm 18$ & $428 \pm 5.5$ & $104 \pm 1.4$ & $28.9 \pm 1.1$ & -2.67 & III or MS & - & \\
J150148.35-630310.1 & $1430 \pm 43$ & $3070 \pm 120$ & $3620 \pm 61$ & $2110 \pm 120$ & $1740 \pm 54$ & $499 \pm 8.3$ & $258 \pm 5.3$ & -2.14 & III or MS & - & \\
J150149.93-630448.2 & $560 \pm 14$ & $1150 \pm 48$ & $1140 \pm 47$ & $580 \pm 6.4$ & $323 \pm 3.5$ & $80.3 \pm 1$ & $25.5 \pm 1$ & -2.63 & III or MS & - & \\
J150153.64-630611.7 & $1230 \pm 30$ & $2360 \pm 98$ & $2550 \pm 55$ & $1440 \pm 110$ & $736 \pm 14$ & $204 \pm 3.1$ & $76.1 \pm 1.7$ & -2.51 & III or MS & - & \\
J150155.78-631332.5 & $502 \pm 8.4$ & $972 \pm 28$ & $990 \pm 21$ & $575 \pm 17$ & $271 \pm 5.3$ & $61.5 \pm 1$ & $18.3 \pm 1$ & -2.71 & III or MS & - & \\
J150156.33-632315.1 & $42.9 \pm 1$ & $118 \pm 2.4$ & $137 \pm 2.5$ & $87.8 \pm 2$ & $55.5 \pm 1$ & $22.2 \pm 0.4$ & $15.9 \pm 0.8$ & -1.93 & III or MS & - & \\
J150156.59-631728.2 & $15.9 \pm 0.5$ & $30.4 \pm 0.8$ & $28.8 \pm 0.5$ & $17.2 \pm 0.4$ & $8.9 \pm 0.2$ & $6.1 \pm 0.2$ & $11.6 \pm 0.9$ & -1.39 & II & -0.13 & \\
J150159.36-631159.8 & $3.2 \pm 0.1$ & $5.5 \pm 0.2$ & $6.9 \pm 0.2$ & $8.8 \pm 0.3$ & $10.9 \pm 0.2$ & $14.1 \pm 0.3$ & $27.7 \pm 1$ & -0.4 & II & -0.51 & \\
J150203.14-630458.3 & $81.8 \pm 1.9$ & $172 \pm 3.5$ & $184 \pm 4$ & $92 \pm 2.1$ & $68.4 \pm 1.3$ & $23.7 \pm 0.4$ & $10.1 \pm 0.6$ & -2.25 & III or MS & - & \\
J150203.82-625442.2 & $1190 \pm 27$ & $2200 \pm 110$ & $2140 \pm 48$ & $1160 \pm 50$ & $572 \pm 11$ & $135 \pm 2.3$ & $41.4 \pm 1.2$ & -2.69 & III or MS & - & \\
J150205.29-630714.5 & $219 \pm 4.7$ & $475 \pm 22$ & $492 \pm 13$ & $255 \pm 5.9$ & $135 \pm 2.6$ & $35.3 \pm 0.7$ & $13.7 \pm 0.8$ & -2.54 & III or MS & - & \\
J150206.96-632047.7 & - & - & - & $0.2 \pm 0.08$ & $1.1 \pm 0.1$ & $4.1 \pm 0.1$ & $135 \pm 3.1$ & 2.45 & I & - & \\
J150208.37-630332.6 & $7.7 \pm 0.3$ & $13.1 \pm 0.7$ & $11.2 \pm 0.5$ & $8.3 \pm 0.2$ & $4.8 \pm 0.1$ & $2 \pm 0.1$ & $14.3 \pm 0.7$ & -0.89 & II & -0.48 & \\
J150212.23-631525.6 & - & $1 \pm 0.1$ & $7.8 \pm 0.2$ & $33.8 \pm 0.7$ & $137 \pm 2.6$ & $372 \pm 6.5$ & $876 \pm 15$ & 1.03 & I & - & IRAS 14580-6303; Flow I\\
J150215.35-632028.3 & $1.4 \pm 0.1$ & $3.5 \pm 0.1$ & $7.1 \pm 0.2$ & $12.6 \pm 0.2$ & $24 \pm 0.5$ & $105 \pm 1.8$ & $307 \pm 5.7$ & 0.62 & I & 0.01 & \\
J150215.97-631749.2 & $536 \pm 15$ & $435 \pm 9.7$ & $316 \pm 6.7$ & $162 \pm 3.5$ & $90.7 \pm 1.7$ & $17.6 \pm 0.4$ & $14.1 \pm 0.9$ & -2.34 & III or MS & - & \\
J150216.19-633738.2 & $1100 \pm 21$ & $1720 \pm 55$ & $1480 \pm 23$ & $768 \pm 27$ & $376 \pm 7.4$ & $82.5 \pm 1.5$ & $23.5 \pm 1.1$ & -2.78 & III or MS & - & \\
J150217.04-633413.5 & $2480 \pm 48$ & $4260 \pm 160$ & $4040 \pm 79$ & $1960 \pm 110$ & $1200 \pm 33$ & $248 \pm 4.6$ & $76.4 \pm 2.1$ & -2.71 & III or MS & - & \\
J150221.83-631301.9 & $195 \pm 4.9$ & $270 \pm 5.8$ & $223 \pm 4.4$ & $140 \pm 2.7$ & $72.8 \pm 1.4$ & $15.7 \pm 0.4$ & $11.4 \pm 1$ & -2.28 & III or MS & - & \\
J150222.43-630458.4 & $175 \pm 5.4$ & $381 \pm 12$ & $370 \pm 9.3$ & $211 \pm 4.7$ & $127 \pm 2.6$ & $39.8 \pm 0.6$ & $16.3 \pm 0.9$ & -2.34 & III or MS & - & \\
J150226.71-630511.6 & $88.7 \pm 2.2$ & $183 \pm 3.9$ & $186 \pm 3.2$ & $106 \pm 2.3$ & $57.4 \pm 1.2$ & $38.7 \pm 0.7$ & $26.1 \pm 0.9$ & -1.85 & III or MS & - & \\
J150227.87-631843.3 & $8.6 \pm 0.2$ & $15.8 \pm 0.5$ & $17.1 \pm 0.4$ & $10.4 \pm 0.1$ & $8.8 \pm 0.1$ & $7.3 \pm 0.1$ & $34.3 \pm 1.1$ & -0.7 & II & -0.32 & \\
J150227.91-630639.2 & $0.6 \pm 0.03$ & $1.3 \pm 0.1$ & $2.3 \pm 0.1$ & $4.5 \pm 0.1$ & $4 \pm 0.1$ & $6.3 \pm 0.1$ & $14.5 \pm 0.8$ & -0.2 & Flat & -0.93 & \\
J150229.62-631738.1 & $1450 \pm 27$ & $3120 \pm 130$ & $4130 \pm 81$ & $2300 \pm 99$ & $2580 \pm 84$ & $686 \pm 11$ & $317 \pm 6.5$ & -2.1 & III or MS & - & \\
J150236.02-632623.8 & $189 \pm 4$ & $395 \pm 16$ & $454 \pm 13$ & $260 \pm 3.9$ & $130 \pm 1.9$ & $31.9 \pm 0.6$ & $12.3 \pm 0.9$ & -2.55 & III or MS & - & \\
J150236.36-632131.1 & - & - & - & $1.8 \pm 0.1$ & $6.5 \pm 0.1$ & $19.9 \pm 0.4$ & $101 \pm 2.2$ & 1.14 & I & - & \\
J150236.87-633836.0 & $664 \pm 15$ & $953 \pm 30$ & $763 \pm 21$ & $420 \pm 5.8$ & $202 \pm 3.2$ & $41.7 \pm 0.7$ & $10.9 \pm 0.9$ & -2.83 & III or MS & - & \\
J150238.02-631900.3 & - & - & - & $0.2 \pm 0.02$ & $1.4 \pm 0.1$ & 0.32u & $30.3 \pm 1.2$ & 1.66 & I & - & very red; [4.6]-[22]=6.60\\
J150238.90-631122.4 & $119 \pm 2.9$ & $81.4 \pm 1.7$ & $54.7 \pm 1.4$ & $28.8 \pm 0.7$ & $19.3 \pm 0.4$ & $69.7 \pm 1.3$ & $246 \pm 6.2$ & -0.35 & II & 0.39 & \\
J150243.24-631112.2 & $98.6 \pm 3.9$ & $198 \pm 7$ & $211 \pm 4.7$ & $148 \pm 3.3$ & $79.3 \pm 1.6$ & $27 \pm 0.6$ & $16 \pm 1.2$ & -2.11 & III or MS & - & \\
J150244.12-631210.0 & $1.8 \pm 0.1$ & $3.1 \pm 0.2$ & $2.6 \pm 0.1$ & $1.7 \pm 0.03$ & $0.9 \pm 0.1$ & $4.8 \pm 0.3$ & $15.3 \pm 1.1$ & -0.24 & Flat & -0.98 & \\
J150245.32-630905.5 & $202 \pm 4.6$ & $432 \pm 21$ & $390 \pm 9.8$ & $222 \pm 4.5$ & $112 \pm 2.2$ & $26 \pm 0.5$ & $10.8 \pm 0.9$ & -2.54 & III or MS & - & \\
J150245.86-625850.4 & $2960 \pm 75$ & $3250 \pm 120$ & $2360 \pm 53$ & $1100 \pm 45$ & $637 \pm 14$ & $124 \pm 2.1$ & $32.9 \pm 1$ & -2.84 & III or MS & - & \\
J150250.84-631252.8 & $1.9 \pm 0.03$ & $2.5 \pm 0.1$ & $2.3 \pm 0.1$ & $1.3 \pm 0.05$ & $0.7 \pm 0.02$ & $3.7 \pm 0.2$ & $8.8 \pm 0.7$ & -0.42 & II & -1.08 & \\
J150254.68-631301.8 & $27.1 \pm 0.6$ & $56 \pm 1.1$ & $51.9 \pm 1$ & $31.4 \pm 0.4$ & $16.3 \pm 0.2$ & $5.9 \pm 0.2$ & $7.4 \pm 0.6$ & -1.84 & III or MS & - & \\
J150259.25-633713.1 & $590 \pm 15$ & $1090 \pm 45$ & $1090 \pm 21$ & $638 \pm 20$ & $296 \pm 5.8$ & $67.1 \pm 1.3$ & $17.7 \pm 1.1$ & -2.77 & III or MS & - & \\
J150303.32-625443.7 & $436 \pm 11$ & $815 \pm 32$ & $820 \pm 20$ & $590 \pm 5.5$ & $276 \pm 3.1$ & $74.5 \pm 1.3$ & $37.4 \pm 1.1$ & -2.33 & III or MS & - & \\
J150305.02-630912.2 & $343 \pm 7.7$ & $740 \pm 28$ & $726 \pm 20$ & $410 \pm 10$ & $205 \pm 3.6$ & $47.3 \pm 0.8$ & $13.5 \pm 0.8$ & -2.71 & III or MS & - & \\
J150309.26-625821.1 & 1.44u & $0.8 \pm 0.2$ & 2.14u & $1.6 \pm 0.1$ & $1.4 \pm 0.01$ & $5.6 \pm 0.2$ & $13.8 \pm 0.9$ & 0.14 & Flat & - & \\
J150309.36-632428.0 & $8.7 \pm 0.2$ & $13.4 \pm 0.2$ & $13.8 \pm 0.3$ & $10.1 \pm 0.3$ & $8.5 \pm 0.2$ & $12.9 \pm 0.3$ & $31.4 \pm 0.8$ & -0.65 & II & -0.35 & MO94-12\\
J150309.47-631640.1 & $191 \pm 4.6$ & $301 \pm 6.5$ & $260 \pm 5.8$ & $159 \pm 2.4$ & $82.8 \pm 1.2$ & $25.1 \pm 0.4$ & $19.4 \pm 0.8$ & -2.11 & III or MS & - & \\
J150310.36-632358.2 & $2080 \pm 39$ & $2490 \pm 72$ & $1970 \pm 33$ & $957 \pm 36$ & $495 \pm 10$ & $102 \pm 1.8$ & $27.7 \pm 1.1$ & -2.83 & III or MS & - & \\
J150311.55-631051.9 & $1540 \pm 39$ & $3180 \pm 180$ & $4060 \pm 79$ & $1750 \pm 120$ & $2080 \pm 130$ & $1000 \pm 13$ & $479 \pm 8.9$ & -1.92 & III or MS & - & IRAS 14590-6259\\
J150313.83-632146.0 & $3.5 \pm 0.1$ & $5.5 \pm 0.1$ & $4.6 \pm 0.2$ & $2.3 \pm 0.1$ & $1.6 \pm 0.04$ & $2.8 \pm 0.2$ & $12.7 \pm 0.9$ & -0.57 & II & -0.84 & \\
J150314.93-631806.6 & $168 \pm 3.8$ & $353 \pm 10$ & $375 \pm 6.9$ & $202 \pm 1.9$ & $133 \pm 1.6$ & $48.3 \pm 0.7$ & $26 \pm 0.9$ & -2.15 & III or MS - & \\
J150323.80-632258.8 & $200.7 \pm 4.5$ & $324.7 \pm 10$ & $495.6 \pm 7.8$ & $\sim$748 & $\sim$1129 & $\sim$2138 & $\sim$4651 & $\sim$0.0 & Flat? & - & C; vdBH65b\\
J150325.16-631122.1 & $79.4 \pm 1.7$ & $83.2 \pm 1.7$ & $77.6 \pm 1.5$ & $86.3 \pm 1.9$ & $86.1 \pm 1.7$ & $224 \pm 3.2$ & $246 \pm 3.1$ & -0.5 & II & 0.58 & \\
J150326.64-632410.9 & $120 \pm 2.5$ & $439 \pm 17$ & $640 \pm 10$ & $352 \pm 9.2$ & $197 \pm 3.8$ & $79.2 \pm 1.4$ & $42.3 \pm 1.6$ & -2.17 & III or MS & - & \\
J150328.03-632335.4 & $3 \pm 0.2$ & $7.4 \pm 0.3$ & $10.9 \pm 0.3$ & $\sim$11 & $\sim$16 & $\sim$23 & $\sim$149 & $\sim$0.1 & Flat? & - & C; vdBH65b region\\
J150328.78-632316.4 & 0.20u & $0.4 \pm 0.1$ & $1.7 \pm 0.2$ & $\sim$3 & $\sim$24 & $\sim$113 & $\sim$686 & $\sim$1.6 & I? & - & C; vdBH65b region\\
J150328.90-632209.0 & $162 \pm 3.6$ & $301 \pm 14$ & $290 \pm 5.6$ & $161 \pm 3.5$ & $85.8 \pm 1.7$ & $19.1 \pm 0.4$ & $16.9 \pm 1.1$ & -2.22 & III or MS & - & \\
J150329.09-630858.6 & $744 \pm 24$ & $1180 \pm 61$ & $1070 \pm 29$ & $582 \pm 19$ & $305 \pm 6.3$ & $61.5 \pm 1.1$ & $16 \pm 0.9$ & -2.81 & III or MS & - & \\
J150332.29-632356.3 & $1.8 \pm 0.1$ & $8.9 \pm 0.4$ & $21.6 \pm 0.4$ & $21.9 \pm 0.6$ & $33.4 \pm 0.7$ & $73.7 \pm 1.4$ & $155 \pm 3.4$ & -0.15 & Flat & 0.0 & \\
J150332.94-630948.1 & $135 \pm 2.9$ & $282 \pm 17$ & $337 \pm 10$ & $207 \pm 4.8$ & $132 \pm 2.7$ & $56.2 \pm 1.1$ & $27.9 \pm 1.1$ & -2.07 & III or MS & - & \\
J150335.29-632250.4 & $0.5 \pm 0.1$ & $2.4 \pm 0.1$ & $4.7 \pm 0.1$ & $6.4 \pm 0.2$ & $15.1 \pm 0.4$ & $33.4 \pm 0.7$ & $89.9 \pm 2.4$ & 0.27 & Flat & -0.40 & \\
J150341.56-625918.8 & $348 \pm 7.5$ & $634 \pm 27$ & $582 \pm 18$ & $301 \pm 5.9$ & $163 \pm 3$ & $39 \pm 0.6$ & $11.7 \pm 0.8$ & -2.68 & III or MS & - & \\
J150342.13-631936.7 & $179 \pm 3.9$ & $540 \pm 19$ & $728 \pm 21$ & $452 \pm 11$ & $247 \pm 4.6$ & $52.5 \pm 0.9$ & $15.5 \pm 0.9$ & -2.65 & III or MS & - & \\
J150343.90-630035.7 & $218 \pm 4.7$ & $385 \pm 19$ & $418 \pm 8.2$ & $224 \pm 5.5$ & $181 \pm 3.9$ & $71.2 \pm 1.3$ & $29.8 \pm 1.1$ & -2.14 & III or MS & - & \\
J150345.69-632341.2 & $4.4 \pm 0.1$ & $9.7 \pm 0.2$ & $12.3 \pm 0.3$ & $10.8 \pm 0.1$ & $12.6 \pm 0.1$ & $28.6 \pm 0.3$ & $43.6 \pm 0.7$ & -0.46 & II & -0.31 & \\
J150348.08-633207.2 & $10.2 \pm 0.3$ & $24 \pm 0.8$ & $30.3 \pm 0.7$ & $23.4 \pm 0.3$ & $29.4 \pm 0.3$ & $34.3 \pm 0.4$ & $144 \pm 2.4$ & -0.33 & II & 0.02 & IRAS 14596-6320; Flow H\\
J150348.39-632632.3 & - & - & - & 0.04u & $0.4 \pm 0.05$ & 0.23u & $44.3 \pm 1.4$ & - & N/A & - & very red; [4.6]-[22]=8.48\\
J150348.40-631435.8 & $11.3 \pm 0.3$ & $16.5 \pm 0.4$ & $15.7 \pm 0.4$ & $13.4 \pm 0.3$ & $11.5 \pm 0.2$ & $15.9 \pm 0.4$ & $25.5 \pm 1.1$ & -0.79 & II & -0.26 & \\
J150348.97-632343.5 & $0.5 \pm 0.1$ & $3.9 \pm 0.1$ & $10.4 \pm 0.2$ & $9.1 \pm 0.2$ & $17.5 \pm 0.4$ & $77 \pm 1.3$ & $183 \pm 3$ & 0.23 & Flat & -0.12 & \\
J150353.05-632554.8 & - & - & - & $0.4 \pm 0.1$ & $0.7 \pm 0.1$ & $6.9 \pm 0.1$ & $30.3 \pm 0.8$ & 1.29 & I & - & \\
J150355.14-633122.6 & $157 \pm 3.5$ & $437 \pm 20$ & $527 \pm 19$ & $293 \pm 5.8$ & $168 \pm 2.6$ & $40.8 \pm 0.8$ & $14.1 \pm 1.1$ & -2.56 & III or MS & - & \\
J150404.59-625739.1 & $711 \pm 23$ & $1150 \pm 33$ & $1030 \pm 19$ & $545 \pm 12$ & $264 \pm 3.9$ & $57.9 \pm 1$ & $14.8 \pm 0.9$ & -2.83 & III or MS & - & \\
J150405.88-632031.8 & $470 \pm 14$ & $947 \pm 34$ & $978 \pm 25$ & $452 \pm 140$ & $290 \pm 5.7$ & $66.4 \pm 1.3$ & $20.8 \pm 1$ & -2.65 & III or MS & - & \\
J150407.64-625835.0 & $7120 \pm 140$ & $9220 \pm 180$ & $7320 \pm 120$ & $3430 \pm 190$ & $2440 \pm 80$ & $370 \pm 5.9$ & $103 \pm 2.4$ & -2.83 & III or MS & - & \\
J150410.07-630535.0 & $1050 \pm 33$ & $1860 \pm 100$ & $1910 \pm 43$ & $1080 \pm 42$ & $510 \pm 10$ & $115 \pm 2.1$ & $32.7 \pm 1.4$ & -2.75 & III or MS & - & \\
J150410.25-625405.5 & $6860 \pm 130$ & $11700 \pm 220$ & $13600 \pm 390$ & $6420 \pm 360$ & $8600 \pm 1000$ & $1390 \pm 23$ & $591 \pm 9.9$ & -2.35 & III or MS & - & \\
J150414.17-630108.0 & $359 \pm 8.1$ & $639 \pm 14$ & $666 \pm 18$ & $375 \pm 4.2$ & $206 \pm 2.8$ & $76.2 \pm 1$ & $51 \pm 1$ & -2.1 & III or MS & - & \\
J150414.73-633634.0 & $3.6 \pm 0.2$ & $9 \pm 0.3$ & $11.2 \pm 0.4$ & $8.8 \pm 0.2$ & $7.6 \pm 0.1$ & $10.6 \pm 0.3$ & $32.9 \pm 0.9$ & -0.54 & II & -0.46 & \\
J150414.83-632425.4 & $299 \pm 6.6$ & $754 \pm 36$ & $914 \pm 14$ & $535 \pm 16$ & $294 \pm 6$ & $67.1 \pm 1.3$ & $18.8 \pm 1.1$ & -2.67 & III or MS & - & \\
J150415.72-633406.7 & $452 \pm 11$ & $813 \pm 45$ & $810 \pm 20$ & $554 \pm 17$ & $256 \pm 5.5$ & $53.7 \pm 1$ & $15.4 \pm 1.1$ & -2.7 & III or MS & - & \\
J150416.50-631210.4 & $8210 \pm 140$ & $9750 \pm 240$ & $7680 \pm 190$ & $3700 \pm 200$ & $3010 \pm 110$ & $389 \pm 6.1$ & $105 \pm 2.4$ & -2.85 & III or MS & - & \\
\enddata
\tablecomments{1 - The WISEPC designation denotes that the extractions and photometry are from the scope of WISE preliminary data release and the "operations coadds."  These coadds were data products created with an early version of the WISE data pipeline.  The coordinate portion of the ID has the format J$hhmmss.ss$$\pm$$ddmmss.s$.\
2 - Fluxes are in milliJansky.  Fluxes are calculated assuming zero-magnitude fluxes of 1594, 1024, 666.7, 309.54, 171.79, 31.676, and 8.3635 Jy for 2MASS J,H, and K, and WISE Bands 1 through 4, respectively.  Upper limit measurements are denoted with a 'u' following the flux.\ 
3 - Infrared luminosity from approximately 1 to 26 $\mu$m, calculated using WISE and 2MASS data for YSOs with 7-band photometry, and d=700 pc.
4 - 'C' denotes sources in crowded regions, where photometry is unreliable by as much as several tenths of a magnitude.  Fluxes for these sources should be considered estimates.  'Very red' sources (\S \ref{sec-veryred}) are noted and an estimate of the [4.6]-[22] color is provided.  Sources in the Cir-MMS and vdBH65b aggregates are noted.  IRAS sources, H$\alpha$ stars discovered by \citet{mo94}, and objects corresponding to a CO outflow (B99) are noted when they correspond to a single, uncrowded WISE detection.}
\label{tab-sources}
\end{deluxetable}

\end{document}